\documentclass[preprint,showpacs,preprintnumbers,amsmath,amssymb]{revtex4}
\usepackage{graphicx}
\usepackage{dcolumn}
\usepackage{bm}
\usepackage{hyperref}

\begin{document}

\title{Nuclear track emulsion in search for the Hoyle-state in dissociation of relativistic ${}^{12}$C nuclei}

\author{D. A. Artemenkov$^a$,
	V. Bradnova$^a$,
	G.I. Britvich$^b$, 
	E. Firu$^c$, 
	M. Haiduc$^c$,\\ 
	V.A. Kalinin$^b$,
	S.P. Kharlamov$^d$,
	N.K. Kornegrutsa$^a$,
	M. Yu Kostin$^b$,\\
	A.V. Maksimov$^b$,
	E. Mitseva$^{a,e}$,
	A. Neagu$^c$,
	V.A. Pikalov$^b$,
	M.K. Polkovnikov$^b$,\\
	V.V. Rusakova$^a$,
	R. Stanoeva$^{e,f}$,
	A.A. Zaitsev$^{a,d}$,
	I.G. Zarubina$^a$,
	P.I. Zarubin$^{a,d}$}
\email{zarubin@lhe.jinr.ru}

\affiliation {
	${}^{a}$Joint Institute for Nuclear Research (JINR), Dubna, Russia\\
	${}^{b}$State Scientific Center ``Institute of High Energy Physics'' (IHEP),Protvino,Russia\\
	${}^{c}$Institute of Space Research, Magurele, Romania\\
	${}^{d}$P.N. Lebedev Physical Institute of the Russian Academy of Sciences (LPI RAS), Moscow, Russia\\
	${}^{e}$Southwestern University, Blagoevgrad, Bulgaria\\
	${}^{f}$Institute for Nuclear Research and Nuclear Energy, Sofia, Bulgaria\\}

 \pacs{21.60.Gx, 25.75.-q, 29.40.Rg}

\maketitle

\section*{Introduction}
\begin{figure}[t]
	\centerline{\includegraphics*[width=0.8\linewidth]{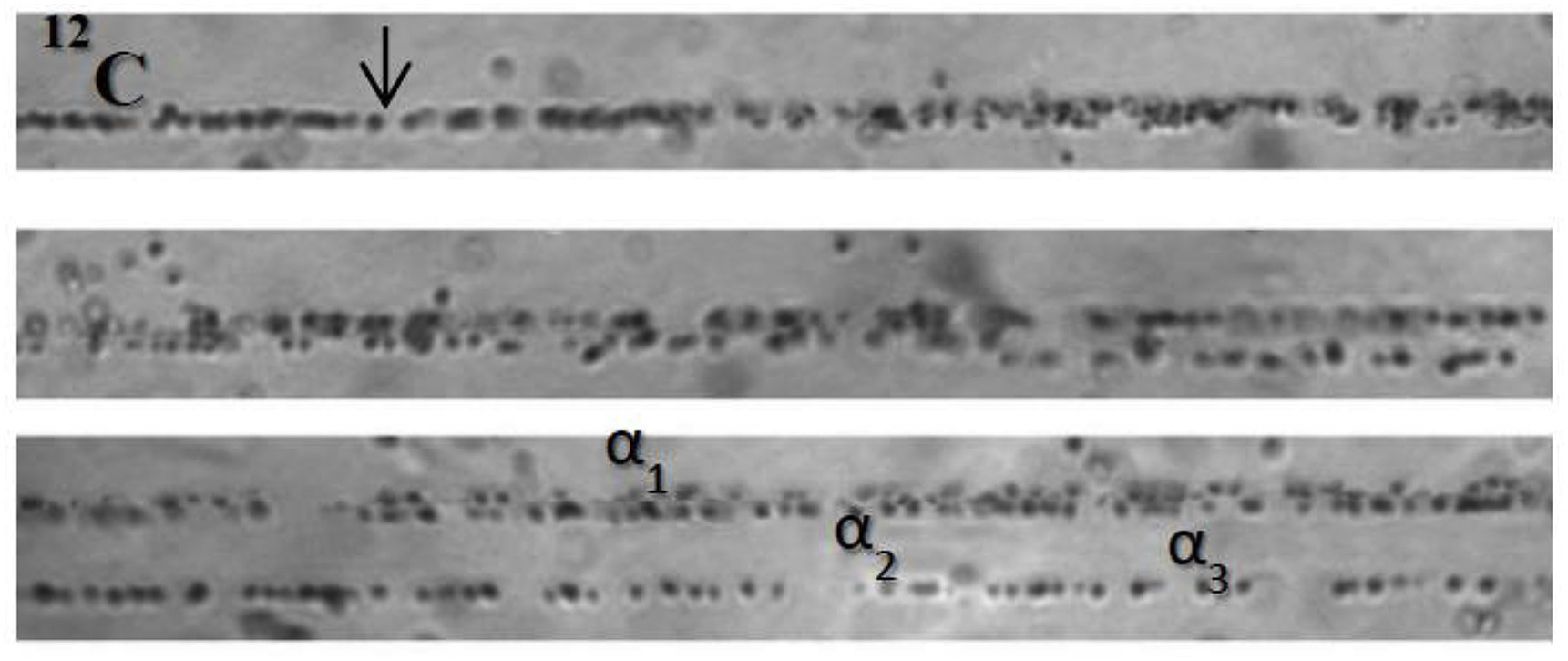}}
	\caption{Consecutive frames of coherent dissociation ${}^{12}$C $\to$ 3$\alpha$ at 1  \textit{A} \, GeV/\textit{c} (``white'' star); arrow indicate interaction vertex; grain sizes are about 0.5 $\mu$m.}
	\label{fig:Fig.1}
\end{figure}

 \noindent Dissociation of relativistic nuclei in a nuclear track emulsion (NTE) is a well-established phenomenon allowing holistic exploration of relativistic ensembles of lightest nuclei. The NTE technique remains the only source of such observations staying unconquerable in sensitivity and angular resolution. Individual features of investigated nuclei manifest themselves in their projectile fragmentation cones. Events of coherent dissociation which does not feature either slow fragments or charged mesons (``white'' stars) are clearly observed in NTE. Since distortions of projectile initial states are minima in them they are especially evident in studies of nuclear structure (example in Fig \ref{fig:Fig.1}) (see Table \ref{tabular:Tab.1}). 
 
 The cluster structure of light nuclei and the role of the unstable ${}^{8}$Be and ${}^{9}$B nuclei in them is a subject of the $\href{http://becquerel.jinr.ru/}{BECQUEREL}$ project (reviewed in \cite{1} and \cite{2}). The studies are performed on a basis of NTE layers longitudinally exposed to relativistic Be, B, C and N isotopes, including radioactive ones. To set new limits for NTE technique it is suggested to search for the Hoyle-state (HS) in dissociation relativistic ${}^{12}$C nuclei using the invariant mass approach. 
 
 Despite capabilities of the NTE technique its history seemed to be completed in the early 2000s. However, since 2012, the company ``Slavich'' (Pereslavl Zalessky, Russia) has resumed production of NTE layers of a thickness from 50 to 200 $\mu$m on a glass base. NTE samples were tested in state-of-art experiments (reviewed in \cite{3}). On the basis of photography on microscopes, the experience of computer recognition of short nuclear tracks in NTE was obtained. At the present time, production of baseless 500 $\mu$m thick layers of is being mastered. Reproduction of NTE allows one to put forward new proposals grounded on this classical technique. 
 
 The status of the experimental and theoretical studies of the second excited state of the ${}^{12}$C nucleus is reviewed in \cite{4}. This excitation is named after the astrophysicist F. Hoyle who postulated its existence to explain the prevalence of the ${}^{12}$C isotope. Following an accurate prediction of the HS energy it was experimentally confirmed that the ${}^{12}$C nucleus has the excited state located at only 378 keV above the mass threshold of the three $\alpha$ particles. Although it is unstable, its width is only 8.5 eV. Such a value indicates that the HS lifetime is comparable with the values for ${}^{8}$Be or $\pi^0$-meson. Observation of HS at a contrast of relativistic energy and the minimum possible energy stored by 3$\alpha$-ensembles can demonstrate HS as a nuclear-molecular object similar to ${}^{8}$Be. First of all it is necessary to establish the very possibility of HS appearance in the relativistic fragmentation cone that is the purpose of the present study.

\begin{table}[t]
	\caption{The average values of $\langle \Theta_{2\alpha} \rangle$ and $\langle \textit{Q}_{2\alpha} \rangle$ ($Q_{2\alpha}$ $<$ 300 keV) in the angular regions of ${}^{8}$Be decays.}
	\begin{center}
		\begin{tabular}{c|c|c} \hline 
			Nucleus ($P_0$, $A$ GeV/$c$) & 
			\begin{tabular}{c}
			$\langle \Theta_{2\alpha} \rangle$ (RMS), $10^{-3}$ rad\\
			 ($Q_{2\alpha}$ $<$ 300 keV)
			\end{tabular} &
			 $\langle \textit{Q}_{2\alpha} \rangle$ (RMS), keV \\ \hline
			${}^{12}$C (4.5) & 2.1 $\pm$ 0.1 (0.8) & 109 $\pm$ 11 (83)\\
			${}^{14}$N (2.9) & 2.9 $\pm$ 0.2 (1.9) & 119.6 $\pm$ 9.5 (72) \\
			${}^{9}$Be (2.0) & 4.4 $\pm$ 0.2 (2.1) & 86 $\pm$ 4 (48) \\
			${}^{10}$C (2.0) & 4.6 $\pm$ 0.2 (1.9) & 63 $\pm$ 7 (83) \\
			${}^{11}$C (2.0) & 4.7 $\pm$ 0.3 (1.9) & 77 $\pm$ 7 (40) \\
			${}^{10}$B (1.6) & 5.9 $\pm$ 0.2 (1.6) & 101 $\pm$ 6 (46) \\
			${}^{12}$C (1.0) & 10.4 $\pm$ 0.6 (3.7) & 117 $\pm$ 12 (75) \\ \hline
		\end{tabular}
		\label{tabular:Tab.1}
	\end{center}
\end{table}

\section*{Reconstruction of invariant mass}

\noindent In general, energy of a few-particle system \textit{Q} \, can be defined as difference between the invariant mass of the system $\textit{M}^*$ \, and a primary nucleus mass or a sum of masses of the particles \textit{M} , that is, $\textit{Q} = \textit{M}^*-M$. \textit{M} is defined as the sum of all products of 4-momenta $\textit{P}_{i,k}$ fragments $\textit{M}^{*2} = (\sum{P}_{j})^2 = \sum (P_{i}P_{k})$. Subtraction of $M$ is a matter of convenience and $Q$ is also named an invariant mass. Reconstruction of $Q$ makes possible to identify decays unstable particles and nuclei.

For the most part, fragments of a relativistic nucleus are contained in a narrow cone of the polar angle $\theta$, which is estimated as $\theta$ = $0.2/P_0$, where the factor 0.2 GeV/$c$ is determined by the spectator-nucleon transverse momentum, while $P_0$ is the momentum of the accelerated projectile nucleon. The fragment 4-momenta $P_{i,k}$ in the cone can be determined in assumption of conservation of momentum per nucleon by fragments of a projectile (or its velocity). This approximation is well grounded when primary energy above 1 $A$ GeV \cite{5}. Then, $Q$ is functionally related with opening angles $\theta$ between fragments. In the ${}^{12}$C context the assumption about the correspondence of a doubly charged fragment to the ${}^{4}$He isotope is well justified also.

The unstable ${}^{8}$Be nucleus is an imminent participant of HS decay and its reconstruction is the precondition of HS identification. The ground state ${}^{8}$Be is sufficiently separated from the first excited state $2^+$ \cite{6} to be identified in a spectrum over the invariant mass $Q_{2\alpha}$ calculated by $\alpha$-pair opening angles $\Theta_{2\alpha}$  \cite{1,2}. Like ${}^{8}$Be, HS is well separated from the higher ${}^{12}$C excitations \cite{6}. Therefore, the same approach can be extended to the identification of HS with respect to the invariant mass of $\alpha$-triples $Q_{3\alpha}$, according to formula\\

$Q_{3\alpha} = \sqrt{\sum\limits_{i\ne{j}} E_{\alpha_{i}}E_{\alpha_{j}} - P_{\alpha_{i}}P_{\alpha_{j}}cos\Theta_{2\alpha}} - 3m_{\alpha}$\\

\noindent ,where $E_{\alpha}$ and $P_{\alpha}$ are energy and momentum values of the $\alpha$-particles $i$ and $j$, $\Theta_{2\alpha}$ is the angle of separation between them, $m_\alpha$ is the mass of the $\alpha$-particle; $P_\alpha = 4P_0$, where $P_0$ is the momentum per nucleon of incident
nuclei.

The foundations of required methods of measurements on microscopes in exposure NTE layers were laid at the beginning of studies on the physics of cosmic rays \cite{7} and, then, used widely beams of relativistic nuclei became available. For these purposes microscopes KSM-1 manufactured by $\mathsf{Carl}$ $\mathsf{Zeiss}$ (Jena) about half of century and still functioning well are applied in JINR. Each microscope is equipped with apochromatic and achromatic lenses providing an increase of 15x and 50x, two eyepieces 12.5x and a tube lens 2x which together give an image magnification 375$-$1250x. The maximum error due to manufacturing tolerances does not exceed 0.05 $\mu$m. Further, samples of 99 events ${}^{10}$B $\to$ 2He + H at 1.6 $A$ GeV/$c$ and 212 events ${}^{11}$C $\to$ 2He + 2H at 2.0 $A$ GeV/$c$ are used to describe in brief the procedure of coordinate measurements as a key aspect of the ongoing HS search.

\begin{figure}[t]
	\centerline{\includegraphics*[width=1\linewidth]{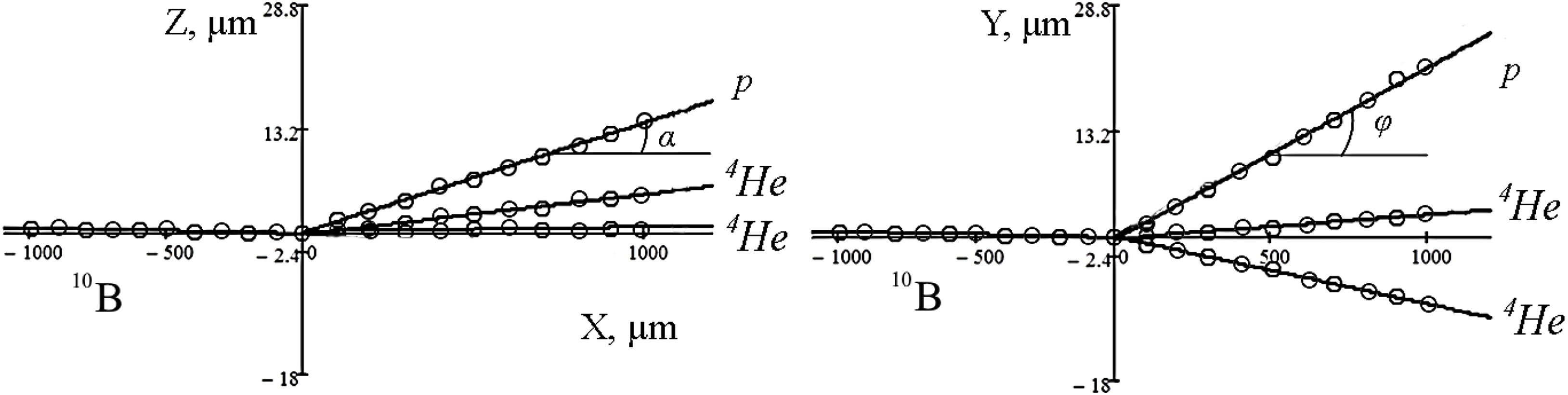}}
	\caption{Example of restored directions in event ${}^{10}$B $\to$ 2He + H over vertical and planar planes.}
	\label{fig:Fig.2}
\end{figure}

\begin{figure}[t]
	\centerline{\includegraphics*[width=0.8\linewidth]{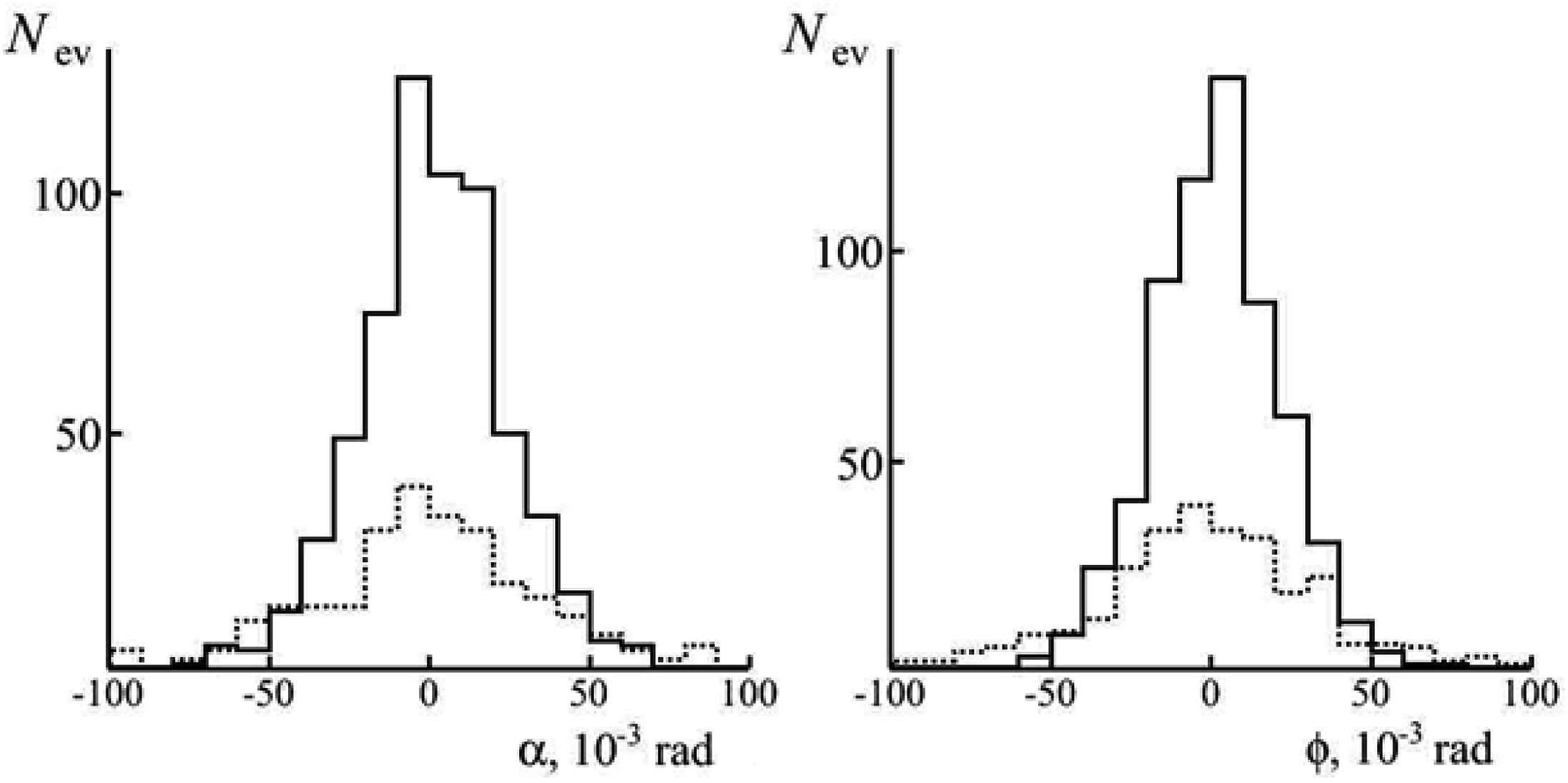}}
	\caption{Distributions of fragments He (solid) and H (dotted) over dip and planar angles $\alpha$ and $\varphi$ in events ${}^{10}$B $\to$ 2He + H.}
	\label{fig:Fig.3}
\end{figure}

When an event in an emulsion plate is found it is fixed on a microscope stage in such a way that direction of a beam track coincides with direction of lengthwise movement of a microscope stage with an accuracy of 0.1 $-$ 0.2 $\mu$m over a 1mm length. All measurements are carried out in 3D-geometry in a Cartesian coordinate system associated with a microscope. Coordinates x, y, z of 10 points at 1mm paths are measured over the primary and fragment tracks. NTE shrinkage due to development has to be taken into account. The found coordinates are linearly approximated to derive the dip and planar angles ($\alpha$ and $\varphi$) for the beam and secondary tracks. The primary track angles are used to transit for the fragment tracks to a coordinate system associated with this track.

\begin{figure}[t]
	\centerline{\includegraphics*[width=0.7\linewidth]{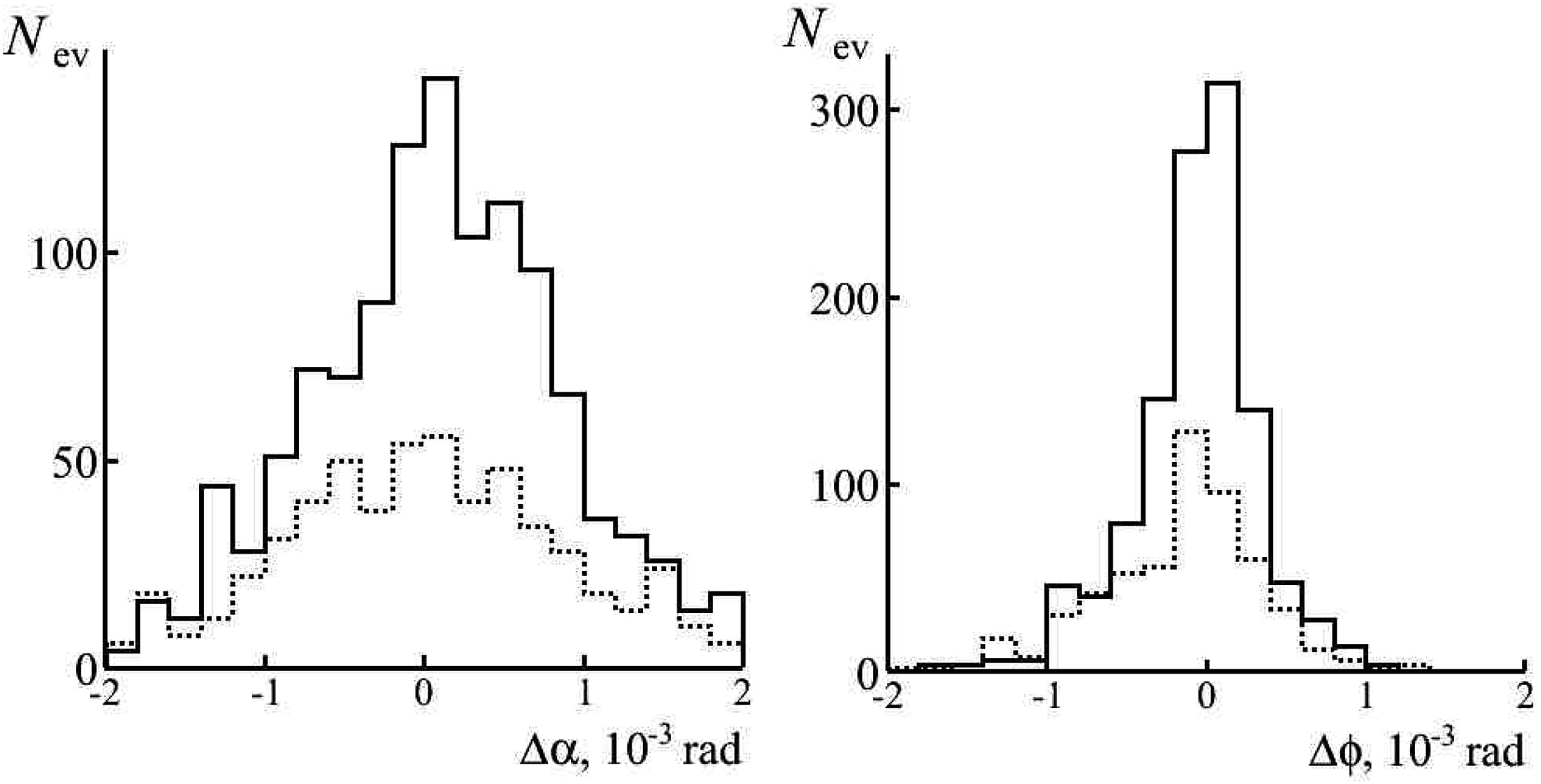}}
	\caption{Distribution of errors in determining dip ($\alpha$) and planar ($\varphi$) angles for fragments He (solid) and H (dotted) in events ${}^{10}$B $\to$ 2He + H.}
	\label{fig:Fig.4}
\end{figure}

\begin{figure}[t]
	\centerline{\includegraphics*[width=0.7\linewidth]{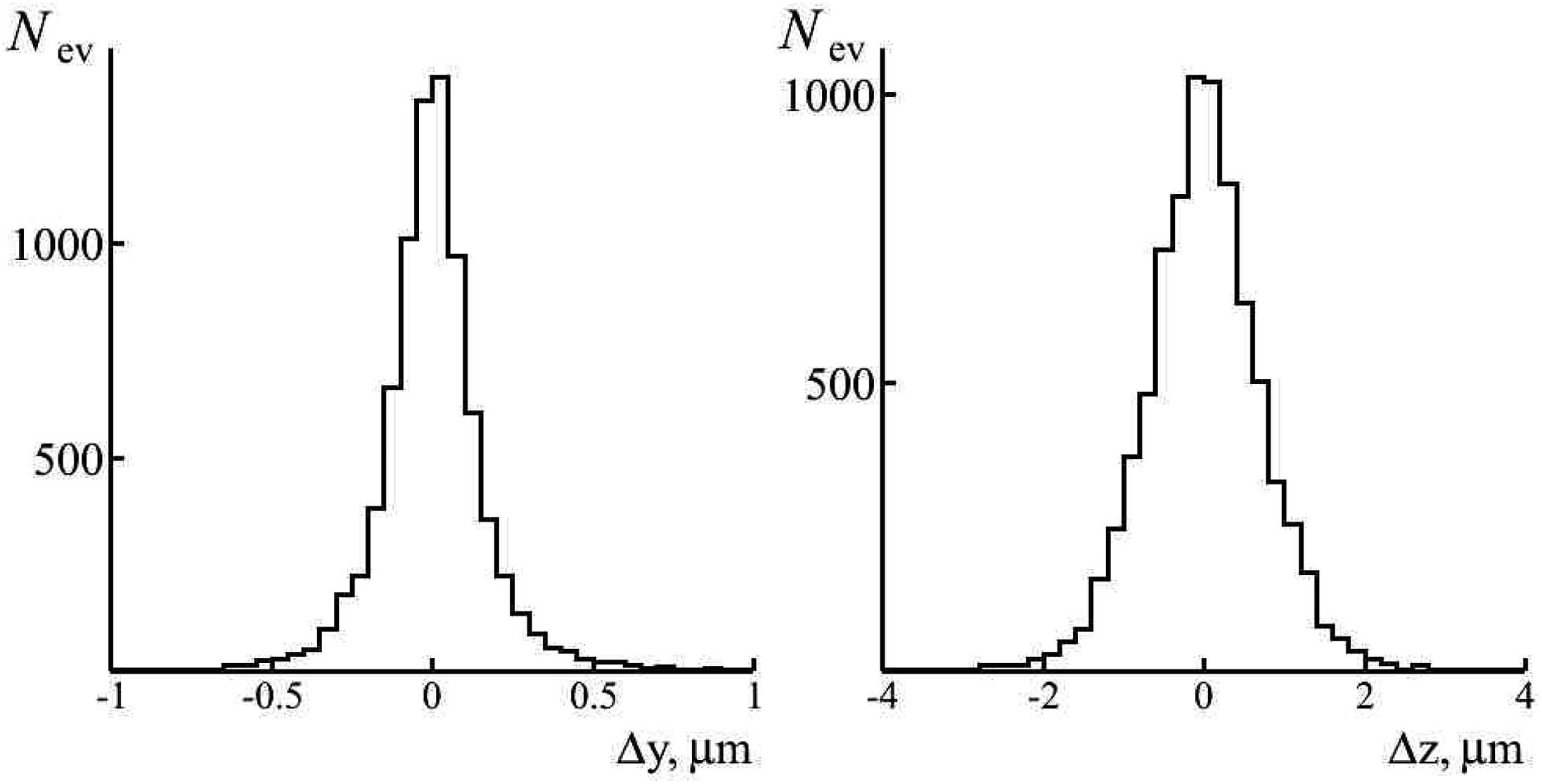}}
	\caption{Distributions of residuals $\Delta$y and $\Delta$z of fitting of coordinates of H and He tracks in events ${}^{10}$B $\to$ 2He + H.}
	\label{fig:Fig.5}
\end{figure}

Fig \ref{fig:Fig.2} shows an outline a reconstructed event ${}^{10}$B $\to$ 2He + H. The distributions of the He and H tracks over the angles $\alpha$ and $\varphi$ are shown in Fig. \ref{fig:Fig.3}. They are characterized by mean values (RMS) $\langle \alpha_{He} \rangle$ = $-$0.7 $\pm$ 0.9 (24) mrad and $\langle \varphi_{He} \rangle$ = 0.2 $\pm$ 0.8 (21) mrad, for the nuclei H – $\langle \alpha_{H} \rangle$ = $-$0.1 $\pm$ 2.0 (37) mrad and $\langle \varphi_{H} \rangle$ = $-$0.7 $\pm$ 2.1 (36) mrad. Mean values (RMS) of angular errors are $\langle \Delta\alpha\rangle$  = 0.08 $\pm$ 0.02 (0.8) mrad and $\langle \Delta\varphi\rangle$ = 0.06 $\pm$ 0.01 (0.4) mrad (Fig. \ref{fig:Fig.4}). Scattering of the coordinate fitting residuals $\Delta$y and $\Delta$z (Fig. \ref{fig:Fig.5}) differs about 4 times and doesn't exceed more than 2 $-$ 3 times a track thickness. A dip coordinate accuracy is less than planar one due to noise generated by vertical displacements of a microscope tube, shrinkage coefficient and vertical NTE distortions during development.

Resulting reconstruction of values $Q_{2\alpha p}$ and $Q_{2\alpha}$ for the ${}^{10}$B and ${}^{11}$C fragmentation is presented in Fig. \ref{fig:Fig.6} in the range which is relevant for ${}^{9}$B ($Q_{2\alpha p}$ $<$ 400 keV). In these cases the ${}^{9}$B decays serves as source of ${}^{8}$Be \cite{2}. Both distributions are similar. Their mean values (at RMS) $\langle Q_{2\alpha p} \rangle$ = 265 $\pm$ 14 (100) keV and $\langle Q_{2\alpha} \rangle$ = 91 $\pm$ 7 (53) keV match the accepted values \cite{6} and expected resolution. Thus, condition $Q_{2\alpha}$ $<$ 200 keV is a practical cut-off for ${}^{8}$Be identification.

\section*{Angular measurements in ${}^{12}$C exposure}

\noindent The current material for the HS search is a set 200 $\mu$m NTE pellicles on 2mm glass of size 9$-$12 cm which is irradiated longitudinally ${}^{12}$C nuclei at initial momentum $P_0$=1 $A$ GeV/$c$. This exposure was performed recently in the medical-biological beam of the Institute of High Energy Physics (Protvino). This ${}^{12}$C beam has energy of about 400 $A$ MeV and used for medical and biological studies. 2\% irradiation homogeneity is provided by application of two rotating electrostatic wobblers. The steps taken in December 2016 and April 2017 resulted in the controllable irradiation with a particle density at the area of irradiation of 2000$-$4500 nuclei/cm$^2$. Accelerated search for 3$\alpha$-events the developed pellicles is carried out by scanning along bands that are transverse to the beam direction. By May 2018, 86 events of ${}^{12}$C $\to$ 3$\alpha$  at 1 $A$ GeV/$c$, including 36 ``white'' stars, are founded and measured following the described procedure.

Besides, measurements made in the 90s in NTE layers exposed to ${}^{12}$C beam at momentum $P_0$ = 4.5 $A$ GeV/$c$ at the JINR
Synchrophasotron are available for 72 (G.M. Chernov's group, Tashkent) \cite{8} and 114 ``white'' stars ${}^{12}$C $\to$ 3$\alpha$ (A.Sh. Gaitinov's group, Alma-Ata) as a legacy of the emulsion community. At that time, the HS problem was not set. Fig. \ref{fig:Fig.4} shows jointly distributions of $\alpha$-particles at both momentum values over the polar emission angle $\theta_{\alpha}$. They are described by the Rayleigh distribution with the parameters $\sigma_{\theta_{\alpha}}$ equal to 27 $\pm$ 3 (1.0 $A$ GeV/$c$) and 6.5 $\pm$ 0.6 (4.5 $A$ GeV/$c$) corresponding to a simple inverse relationship between $P_0$ and $\sigma_{\theta_{\alpha}}$. In addition, Fig. \ref{fig:Fig.7} shows data on He fragments for the 2.0 $A$ GeV/$c$ ${}^{11}$C dissociation where the ${}^{4}$He isotope dominates.

\begin{figure}[t]
	\centerline{\includegraphics*[width=0.8\linewidth]{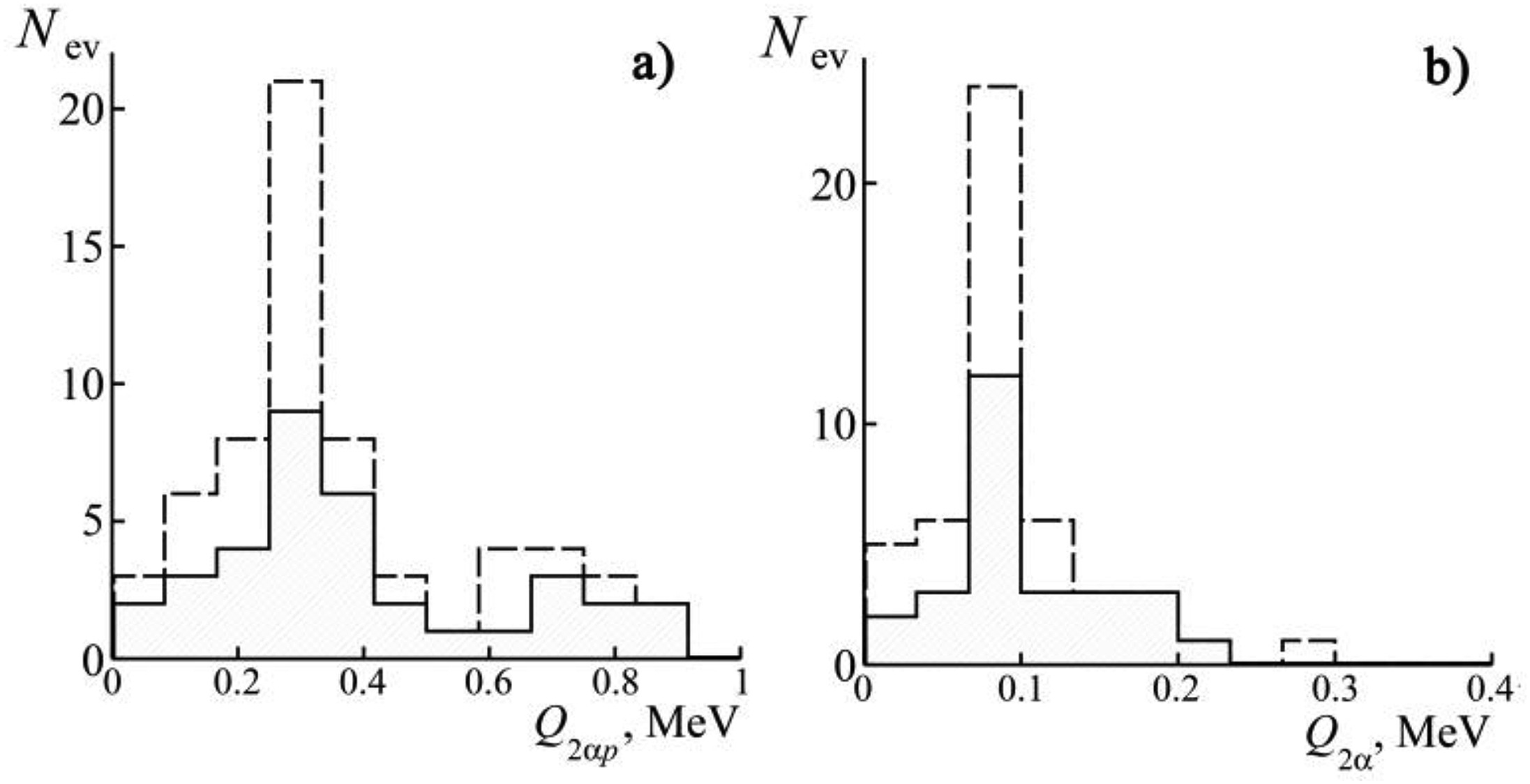}}
	\caption{Distributions of triples 2$\alpha p$ over invariant mass $Q_{2\alpha p}$ (a) for fragmentation ${}^{10}$B $\to$ 2He + H at 1.6 $A$ GeV/$c$ (solid) and ${}^{11}$C $\to$ 2He + 2H at 2.0 $A$ GeV/$c$ (added,dashed) and $Q_{2\alpha}$ of $\alpha$-pairs in ${}^{9}$B decays identified in these events (b).}
	\label{fig:Fig.6}
\end{figure}

\begin{figure}[t]
	\centerline{\includegraphics*[width=0.45\linewidth]{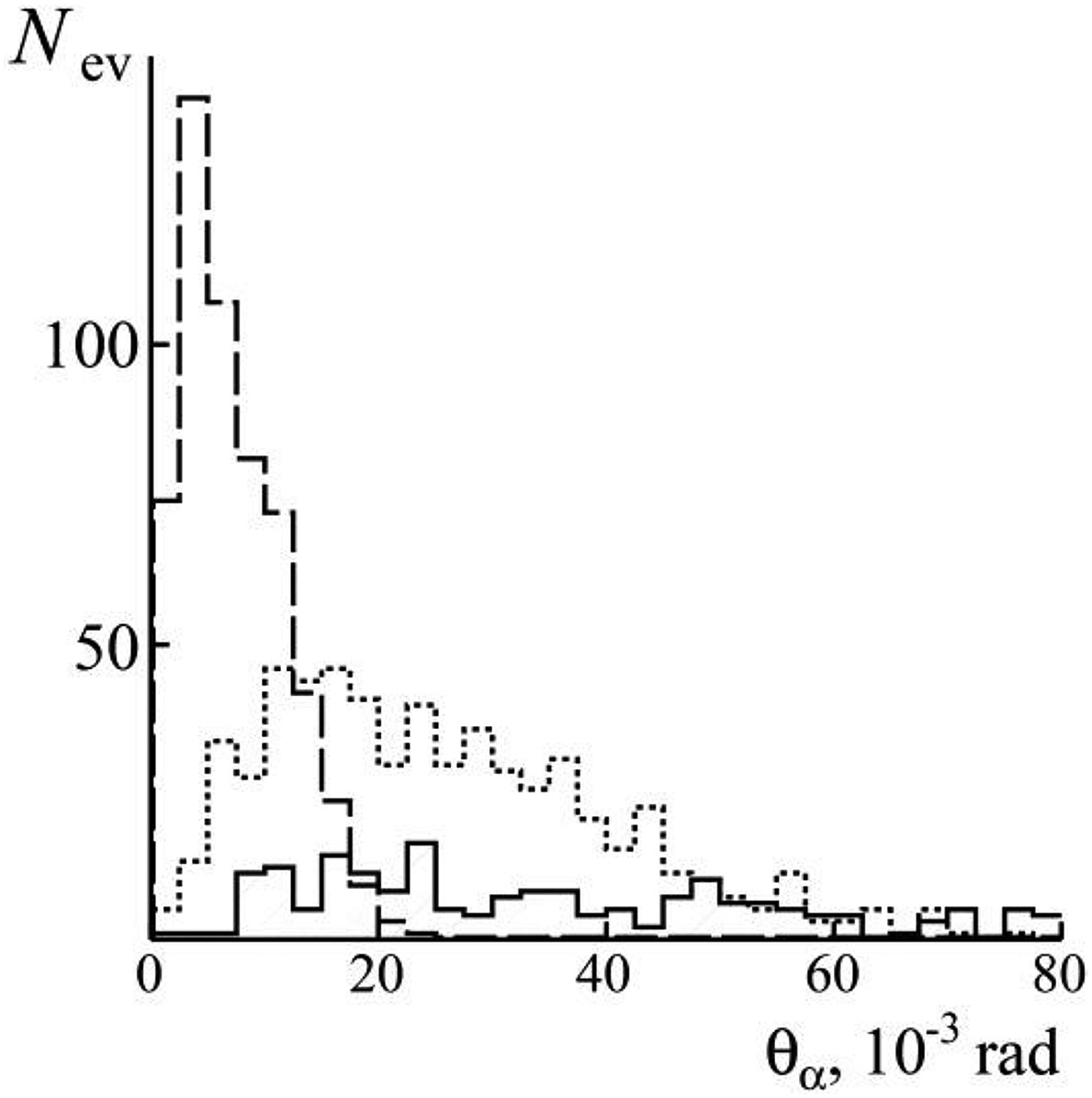}}
	\caption{Distribution over polar angle $\theta_\alpha$ of relativistic He fragments in exposures at 4.5 (dashed) and 1 $A$ GeV/$c$ (solid) ${}^{12}$C and 2.0 $A$ GeV/$c$ ${}^{11}$C (dotted).}
	\label{fig:Fig.7}
\end{figure}

\begin{figure}[t]
	\centerline{\includegraphics*[width=0.6\linewidth]{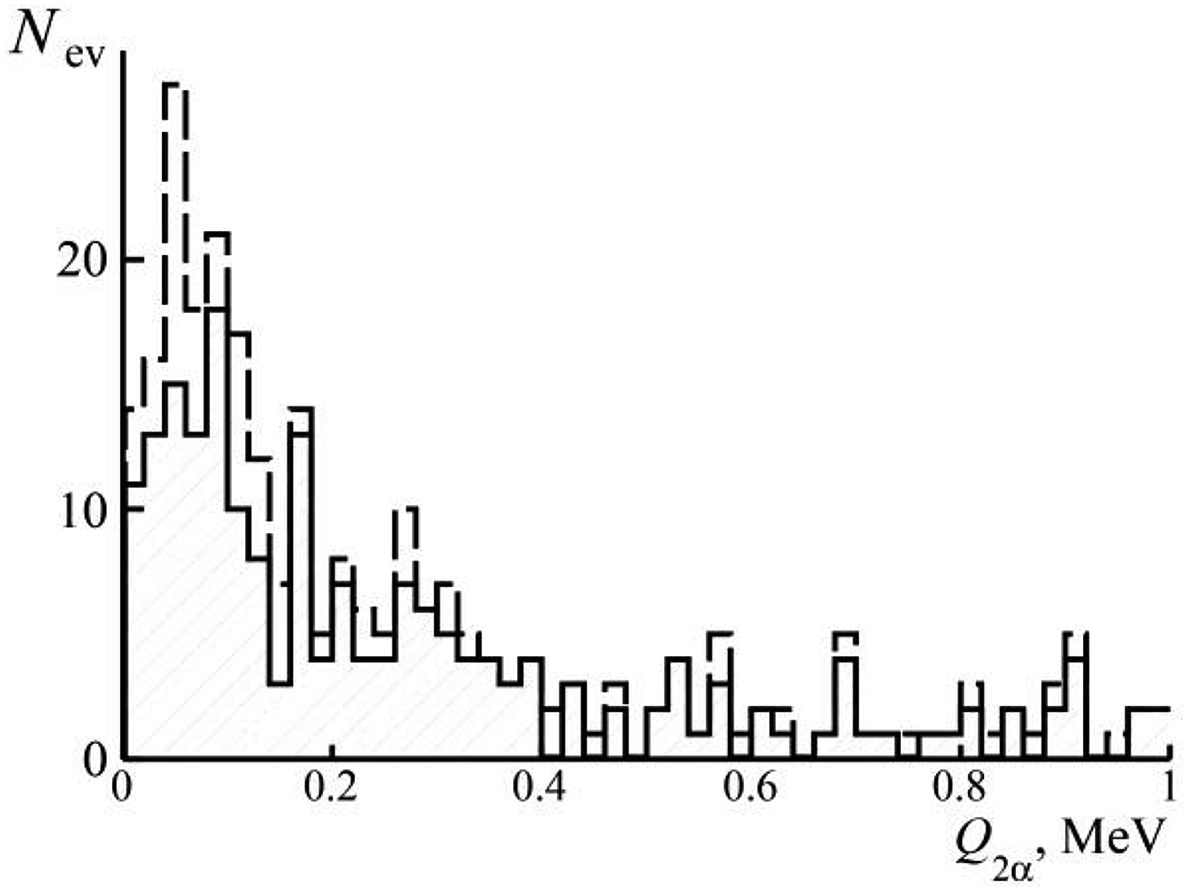}}
	\caption{Distribution of $\alpha$-pairs over invariant mass $Q_{2\alpha}$ $<$ 1 MeV in the dissociation ${}^{12}$C →$\to$ 3$\alpha$ at 4.5 (solid) and 1 $A$ GeV/$c$ (added, dashed).}
	\label{fig:Fig.8}
\end{figure}

\begin{figure}[t]
	\centerline{\includegraphics*[width=0.5\linewidth]{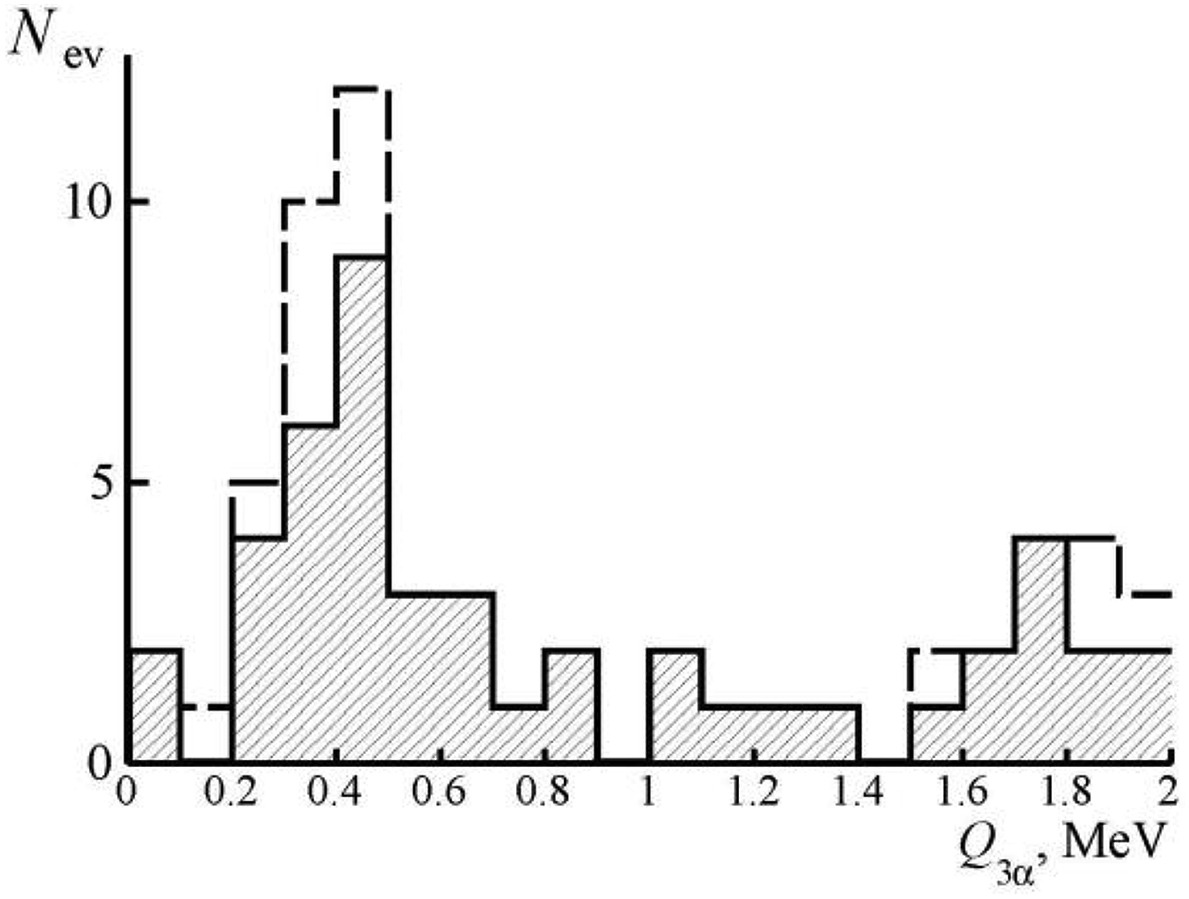}}
	\caption{Distribution of $\alpha$-triples over invariant mass $Q_{3\alpha}$ $<$ 2 MeV in dissociation of ${}^{12}$C →$\to$ 3$\alpha$ at 4.5 $A$ GeV/$c$ (solid) and 1 $A$ GeV/$c$ (added, dashed).}
	\label{fig:Fig.9}
\end{figure}

The $Q_{2\alpha}$ distributions obtained on a basis of angular measurements of events ${}^{12}$C $\to$ 3$\alpha$ at two values $P_0$ are presented jointly in Fig. \ref{fig:Fig.8}. Both are distributions do not differ within statistics. The region $Q_{2\alpha}$ $<$ 200 keV contains a peak pressed to the origin which corresponds to decays of ${}^{8}$Be. Although the ${}^{8}$Be signal is present the $Q_{2\alpha}$ distribution appears to be significantly wider than in Fig. \ref{fig:Fig.6}(b).

In the $Q_{3\alpha}$ distribution over the invariant mass of the $\alpha$-triples (Fig. \ref{fig:Fig.9}) there is a peak in the region $Q_{3\alpha}$ $<$ 1 MeV where HS decays could be reflected. For events at 4.5 $A$ GeV/$c$ the mean value for the events at the peak $\langle Q_{3\alpha} \rangle$ (at RMS) is 441 $\pm$ 34 (190) keV, and at 1 $A$ GeV/$c$, respectively, 346 $\pm$ 28 (85) keV. According to the ``soft'' condition $Q_{3\alpha}$ $<$ 1 MeV in the 4.5 $A$ GeV/$c$ exposure 30 (of 186) events can be attributed to HS and 9 (of 86) including 5 ``white'' stars (of 36) in 1 $A$ GeV/$c$ exposure.

\begin{figure}[t]
	\centerline{\includegraphics*[width=0.45\linewidth]{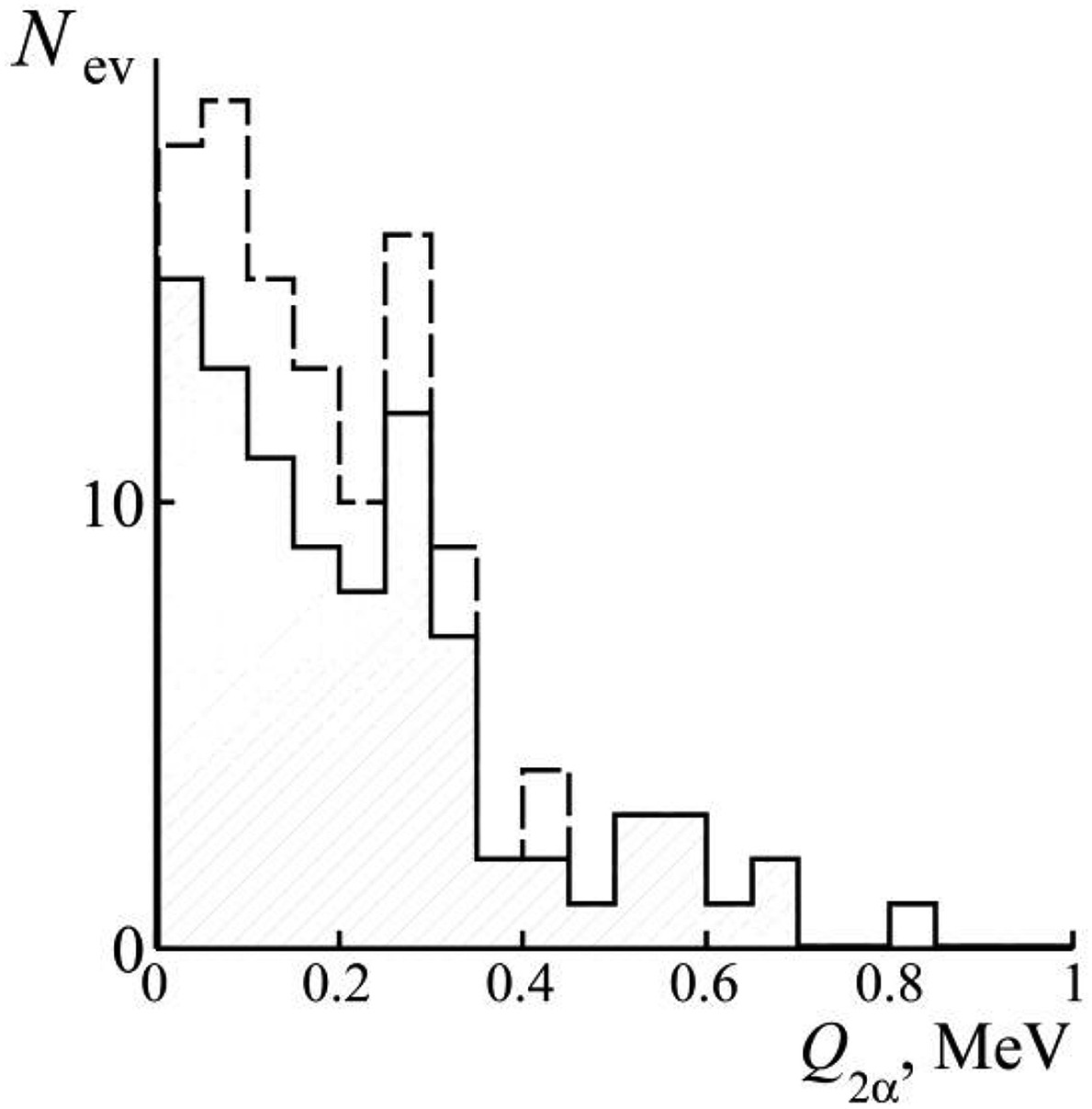}}
	\caption{Distribution of $\alpha$-pairs over invariant mass $Q_{2\alpha}$ in the HS like decays ($Q_{3\alpha}$ $<$ 1 MeV) in dissociation of ${}^{12}$C →$\to$ 3$\alpha$ at 4.5 (solid) and 1 $A$ GeV/$c$ (added, dashed).}
	\label{fig:Fig.10}
\end{figure}

\begin{figure}[t]
	\centerline{\includegraphics*[width=0.6\linewidth]{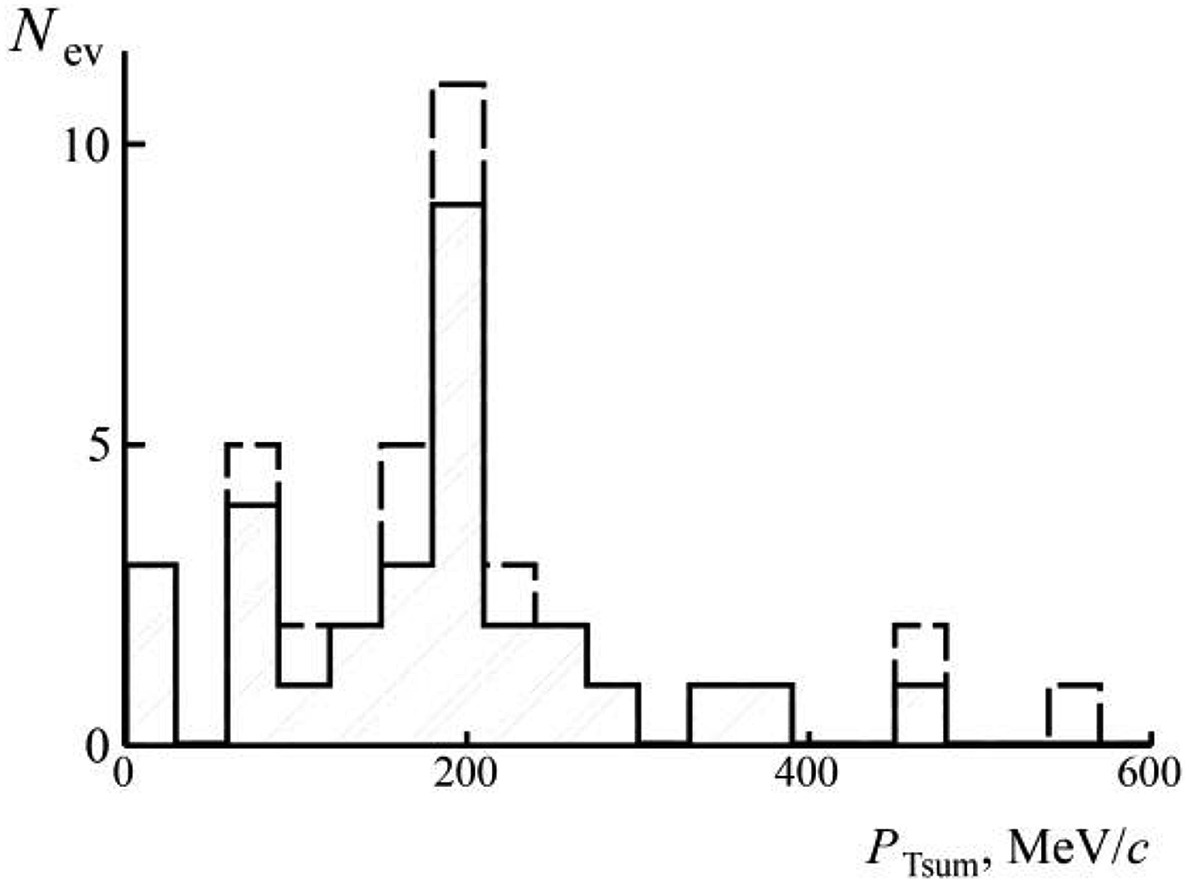}}
	\caption{Distribution of $\alpha$-triples of HS like decays ($Q_{3\alpha}$ $<$ 1 MeV) over total transverse momentum $\langle P_{Tsum} \rangle$ in dissociation ${}^{12}$C →$\to$ 3$\alpha$ at 4.5 (solid) and 1 $A$ GeV/$c$ (added, dashed).}
	\label{fig:Fig.11}
\end{figure}

When selecting $\alpha$-pairs from $\alpha$-triples that correspond to the HS criterion $Q_{3\alpha}$ $<$ 1 MeV the $Q_{2\alpha}$ distribution acquires the form shown in Fig. \ref{fig:Fig.10}. The average value $\langle Q_{2\alpha} \rangle$ (RMS) is 210 $\pm$ 15 (156) keV. The distribution form becomes wider and separation of the ${}^{8}$Be peak in the region $Q_{2\alpha}$ $<$ 200 keV is impossible. This change is caused by the increased contribution of non-${}^{8}$Be-resonance $\alpha$ pairs of HS decays masking the ${}^{8}$Be signal. In turn, this circumstance makes unattainable a more detailed analysis of the HS inner structure. It characterizes a limitation of our approach to penetrate in the HS structure. Nevertheless, it is concluded that HS is observed in a relativistic dissociation ${}^{12}$C $\to$ 3$\alpha$ with probability about 10-15\%.

The angular measurements make it possible to conclude about the dynamics of the HS appearance according to the distribution of $\alpha$-particle triples over their total transverse momentum $P_{Tsum}$ (Fig. \ref{fig:Fig.11}). Its average value $\langle P_{Tsum} \rangle$ (RMS) is equal to 190 $\pm$ 19 (118) MeV/$c$ corresponding to the nuclear-diffraction mechanism. In the case of electromagnetic dissociation on Ag and Br nuclei composing NTE the limitation is expected to be $P_{Tsum}$ $<$ 100 MeV/$c$  \cite{9}. It is surprising that such a ``fragile'' formation of three $\alpha$-particles as HS can arise in relativistic collisions as an ensemble which is ``bouncing off'' with the transverse momentum $P_{Tsum}$ characteristic for strong interactions rather than electromagnetic ones. It can be assumed that increased statistics allow registration of the HS formation outside the angular cone of fragmentation of the parent nucleus. Events of this kind were observed in the cases ${}^{9}$Be $\to$ ${}^{8}$Be and ${}^{10}$C $\to$ ${}^{9}$B. Such observations would clearly demonstrate HS as a holistic and long-lived nuclear-molecular state.

\section*{Conclusion}

\noindent In dissociation ${}^{12}$C $\to$ 3$\alpha$ at 4.5 and 1 $A$ GeV/$c$ in nuclear track emulsion production of the Hoyle's-state is identified by using approximate invariant mass representation. Contribution of HS is estimated to be about 10-15\%. This conclusion is grounded on the basis of the most precise angular measurements performed by three research groups in two exposures at two momentum values that are separated in time by two decades. By itself, this finding demonstrates the thoroughness of the NTE technique.

However, the NTE grounded approach doesn't allow one to address the features of the HS decay. Nevertheless reconstruction of HS in NTE by the invariant mass of relativistic $\alpha$-triples can be applied to study processes with the HS formation as a wholesome relativistic object at large moment transfers. It is possible that a HS wouldn't be limited only as the ${}^{12}$C excitation but can manifest itself similarly to ${}^{8}$Be as a universal object in fragmentation of heavier nuclei. In this respect, the closest source to verify such a assumption is the ${}^{14}$N nucleus. The ${}^{13}$N and ${}^{13}$C nuclei whose beams can be formed in the ${}^{14}$N fragmentation are even more convenient in this respect.

\end {document}